\documentclass[aps,pra,twocolumn,twoside,a4paper,showpacs,floatfix]{revtex4}

\usepackage{graphicx}
\usepackage{amsmath,amsthm,amsfonts}
\usepackage{psfrag}
\usepackage{fullpage}
\usepackage{textcomp}
\usepackage{tabularx}
\usepackage{booktabs}
\usepackage{array}
\usepackage{bbm}
\usepackage{bm}
\usepackage{xcolor}
\usepackage{enumerate}

\newcommand{\comment}[1]{}
\newcommand{\hide}[1]{}

\renewcommand{\phi}{\varphi}

\renewcommand\appendix{\par
  \setcounter{section}{0}
  \setcounter{subsection}{0}
  \setcounter{figure}{0}
  \setcounter{table}{0}
  \renewcommand\thesection{Appendix \Alph{section}}
  \renewcommand\thefigure{\Alph{section}\arabic{figure}}
  \renewcommand\thetable{\Alph{section}\arabic{table}}
}

\usepackage{epstopdf}
\DeclareGraphicsExtensions{ .png,.PNG,.pdf,.PDF, .jpg,.mps,.jpeg,.jbig2,.jb2,.JPG,.JPEG,.JBIG2,.JB2}

\begin{document}

\title{Logical and inequality based contextuality for qudits}
\author{Adel Sohbi} \email{adel.sohbi@telecom-paristech.fr}
\author{Isabelle Zaquine}
\author{Eleni Diamanti}
\author{Damian Markham}
\affiliation{LTCI, CNRS, Telecom ParisTech, Univ Paris-Saclay, 75013 Paris, France}

\begin{abstract}
In this work we present a generalization of the recently developed Hardy-like logical proof of contextuality and of the so-called KCBS contextuality inequality for any qudit of dimension greater than three. Our approach uses compatibility graphs that can only be satisfied by qudits. We find a construction for states and measurements that satisfy these graphs and demonstrate both logical and inequality based contextuality for qudits. Interestingly, the quantum violation of the inequality is constant as dimension increases. We also discuss the issue of imprecision in experimental implementations of contextuality tests and a way of addressing this problem using the notion of ontological faithfulness.
\end{abstract}


\date{\today}
\maketitle

\section{Introduction}

Contextuality is one of the key features of quantum mechanics, of which non-locality can be considered as a special case. The underlying question is whether or not it is possible to assign values to measurement outcomes, independent of which context they are measured in. This implicitly involves the notion of compatibility, which refers to whether measurements can be performed at the same time, or in sequence, without affecting each other. A context is then given by the set of measurements that can be performed together - or that are compatible. In quantum mechanics, for instance, two projective measurements are compatible if they commute. Several general frameworks have emerged in recent years to describe non-locality and more broadly contextuality \cite{AB:njp11,FLS:qpl13,CSW:arXiv10}. Both notions have also become recognized as resources in quantum information. Non-locality has many powerful applications such as in communication complexity \cite{BZP:prl04}, randomness amplification \cite{RBG:arxiv13}, and device independence \cite{ABG:prl07,BGL:prl11}, while more general contextuality has been increasingly identified as key behind quantum computational power \cite{AB09,raussendorf2013contextuality,HWVE:nat14} and cryptographic applications \cite{SBKTP09,HHHHPB10}.

Identifying methods for witnessing these fundamental notions is essential for their study and applicability. For non-locality the most commonly used method is the Bell-like inequalities, which look at statistics of measurements over separated parties to judge whether these can be considered to result from a local hidden variable (LHV) model. Similar inequalities exist for contextuality - where from statistics one can judge whether the theory has a non-contextual description. A prominent example is the so-called KCBS inequality \cite{KCBS:prl08}. Another means to expose contextuality and non-locality is through logical contradictions with the existence of non-contextual or LHV models, respectively. The first famous example of this for non-locality was presented as a paradox by Hardy \cite{Hardy:prl93}; there if certain events happen and certain others are excluded, LHV implies that some events may or may not be possible, in contradiction with quantum mechanics. In \cite{CBCB:prl13} this approach was extended to contextuality, in the case of qutrits (quantum systems spanning a three-dimensional Hilbert space). Both of these approaches can be described using the general frameworks of \cite{AB:njp11,FLS:qpl13,CSW:arXiv10}.

Contextuality has been observed experimentally in various physical systems over the past few years \cite{ARBC:prl09,KZGKGCBR:nat09,ZWDCLHYD:prl12,ADTPBC:prl12,AHANBSC:prx13,BCAFACCP:pra14,BJMC:prl14,ACGBXLABSC:pra15} involving a variety of tests. Typically, to perform these tests it is necessary to encode the information on several degrees of freedom of single photons. Despite these advances, the experimental characterization of contextuality is still a subject of controversy \cite{BK04,Spekkens05,Spekkens14,Winter:jpa14} due to the existence of loopholes in practical realizations; for instance, one crucial problem lies in being sure that the same measurement genuinely appears in different contexts stemming from experimental imprecisions.

In this paper we are interested in the problem of witnessing contextuality for qudits spanning a Hilbert space of dimension greater than three. We study both aforementioned methods, and provide an extension of the Hardy-like contextuality test as well as a proof of the violation of an extended KCBS inequality. To this end, we use the framework of \cite{CSW:arXiv10}, which was also used in \cite{CBCB:prl13}. Our extension is constructive and requires qudits to satisfy the necessary compatibility relations. Interestingly, we find that the quantum violation of the inequality remains constant as dimension increases. We finally discuss issues arising from imprecisions in experimental implementations and suggest an approach to taking them into account using the notion of ontological faithfulness introduced in \cite{Winter:jpa14} and applying it to our results.

The paper is structured as follows. In Section II, we present the preliminary notions; we introduce how a graph can be used to represent measurement contexts and we recall the KCBS inequality and the Hardy-like proof of contextuality. In Section III, starting from the pentagon graph proposed in \cite{CBCB:prl13} we present our graph construction for higher dimensions, and then provide the Hardy-like proof of contextuality and the extension of the KCBS inequality. We also give a set of measurements and a qudit state that lead to contextuality for both tests and describe a way of visualizing them using the so-called Majorana representation. Finally, in Section IV, we discuss implementation issues in contextuality tests.

\section{Preliminary notions}
We first review the graphical formalism of \cite{CSW:arXiv10} (see also \cite{CBCB:prl13,Winter:jpa14}).
We define a graph $\mathcal{G}(V,E)$ of $N$ vertices, for which we associate to each vertex $i\in V$ a dichotomic measurement outcome $X_i=0$ (`no') or $X_i = 1$ (`yes') and where the edges represent the exclusivity and the compatibility of the measurements. When querying the value of $X_i$ we say we are \emph{measuring} $X_i$. Measurements are compatible if it is possible to perform them simultaneously. Dichotomic measurements are exclusive if they cannot both have an output `yes', $i.e.$ it is not possible that exclusive measurements have the outcome $1$ simultaneously. Thus for all adjacent vertices $i\in V$ and $j\in V$, the probability to have the measurement outcome 1 assigned to both vertices is:
\begin{equation}\label{eq:hp1}
P(1,1\vert i,j)=0,
\end{equation}
where $p(a,b...|c,d...)$ represents the probability of getting results $a,b...$ given measurement settings $c,d...$.

If a vertex $i$ has two neighbors $j$ and $k$, then $X_i$ can be measured with $X_j$ or with $X_k$. We call the choice of a pair (or more generally of a set) a context, $C$ (denoting the set of vertices jointly measured). If $j$ and $k$ are not connected, then they cannot be measured at the same time and therefore $X_j$ and $X_k$ correspond to two different contexts for the measurement of $X_i$.

We now consider how different classes of physical theories can assign outcomes on a given graph.

In a deterministic non-contextual model, each dichotomic measurement leads to a predefined outcome $0$ or $1$. Hence the outcome is independent of the measurement context. In this way each vertex $i\in V$ of the graph $\mathcal{G}(V,E)$ has an assigned value $X_i$ corresponding to a measurement outcome. As explained before, because the edges of the graph represent the exclusivity of the measurements, two adjoint vertices on the graph cannot simultaneously have the outcome value $1$ assigned. A general (probabilistic) non-contextual hidden variable model is one where the choice of deterministic assignment can be made according to some probability distribution.

In quantum physics, the dichotomic measurements we will use will be represented by rank one projectors $\{P_i=\vert v_i \rangle \langle v_i \vert\}$ with the normalized eigenvectors $\vert v_i \rangle$, where outcome $X_i=1$ is associated to projector $P_i$ and outcome $X_i=0$ is associated to $I-P_i$. This is equivalent to associating a unit vector $\vert v_i \rangle$ to the vertex $i$. In this framework, the exclusivity and compatibility relations between two measurements correspond to an orthogonality relation between the two unit vectors of the adjacent vertices. This is called an orthonormal representation of a graph. In this way a complete subgraph (\emph{i.e.} a set of vertices that are all connected to each other, known as \textit{clique}), corresponds to a set of mutually orthogonal states. This implies that the dimension $d$ of the quantum system depends on the connectivity of the graph: $d$ must be at least as big as the size of the largest complete subgraph (the maximal clique).

\subsection{KCBS inequality}

The graphs defined above can be used to derive non-contextuality inequalities as follows. For a given graph, a complete subgraph represents a compatible set of measurements - or context $C$. 
The exclusivity condition implies that $\sum_{i\in C} X_i \leq 1$. For classically assigned $X_i$ we then arrive at the following inequality (see e.g. \cite{CBCB:prl13,Winter:jpa14}),
\begin{equation}\label{eq:generalkcbs}
\sum_{i\in V}  \langle X_i \rangle\leq \alpha(G),
\end{equation}
where $\alpha(G)$ is the independence number of the graph $G$ (\emph{i.e.} the maximum number of vertices that are not connected to each other) and $\langle X_i \rangle$ is the expectation of the value of $X_i$. Note here that a set is independent if and only if it is a clique in the graph's complement, so the two notions are complementary.

\begin{figure}[tb]
      \centering
      \includegraphics[width=4cm]{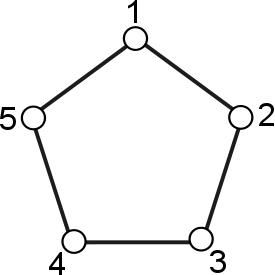}
      \caption{The pentagon compatibility graph corresponding to the KBCS  inequality \cite{KCBS:prl08} and Hardy-like paradox introduced in \cite{CBCB:prl13}.
It can easily be seen that $\alpha(G)=2$.}
			\label{fig:N5}
\end{figure}

For the pentagon graph depicted in Fig.~\ref{fig:N5}, Eq.~(\ref{eq:generalkcbs}) becomes the KCBS inequality \cite{KCBS:prl08},
\begin{equation}\label{eq:kcbs}
\sum_{i=1}^5 \langle X_i \rangle  \leq 2.
\end{equation}

The quantum violation of this inequality is given by $\sum\langle X_i \rangle = \sum|\langle v_i|\psi \rangle |^2$, the maximum of which is known as the Lovasz function of the graph, denoted $\vartheta(G)$. Thus for a graph $G$ we can define the extended KCBS inequality \cite{CBCB:prl13},
\begin{align}\label{eq:beta}
\beta = \sum_{i\in V} \langle X_i \rangle \leq \alpha(G)\leq \vartheta(G).
\end{align}
For the pentagon graph, the maximum quantum violation is $\vartheta(G) = \sqrt{5}$ \cite{KCBS:prl08}.

\subsection{Hardy-like paradox}

By imposing additional conditions on the outcome probabilities to those imposed by the graph itself, we can arrive at Hardy-like logical contradictions with non-contextual hidden variable models.

We start with the example of the pentagon. 
This is a cyclic graph, hence the exclusivity relations can be written as:
\begin{equation}\label{eq:penta}
P(1,1\vert i,i+1)=0.
\end{equation}

To construct a Hardy paradox we impose the additional two followings conditions:
\begin{align}\label{eq:hp}
P(0,0\vert 2,3)=0,\notag\\
P(0,0\vert 4,5)=0.
\end{align}

We can then easily see that a system that has a deterministic non-contextual description satisfying Eqs.~(\ref{eq:penta}) and~(\ref{eq:hp}) has $P(1\vert 1)=0$. Thus the case where a system verifies Eqs.~(\ref{eq:penta}) and~(\ref{eq:hp}) but has $P(1\vert 1) > 0$, exhibits a contextual description. In this way a logical based proof of contextuality can be built - in quantum physics, it is possible to find a set of measurement vectors $\{ | v_i \rangle \}$ and a state such that Eqs.~(\ref{eq:penta}) and (\ref{eq:hp}) are satisfied and yet $P(1\vert 1)=\frac{1}{9}>0$ \cite{CBCB:prl13}.

We emphasize again here that when constructing the Hardy-like paradox we have more conditions than those given just by the graph. In this way the optimal inequality violation for the graph may not be reached whilst at the same time satisfying these extra conditions. In the example of the pentagon, for instance, the reduced bound is $2 + \frac{1}{9}$ \cite{CBCB:prl13}. Conversely, the state and measurements that violate maximally the inequality are in general not the same as those satisfying the Hardy-like paradox.

\section{Main Results}

\subsection{Building the graph}

We begin our analysis by introducing a family of graphs with $N > 5$ vertices, which generalizes the pentagon example in a way that requires qudits ($d>3$) to demonstrate contextuality.
We start with vertex $1$, which  is connected to all the remaining vertices except vertices $2$ and $N$. Then vertices $2$ and $N$ are connected to each other. Next we define two subsets of the set of vertices $V$: $V_\mathcal{A}=\{2,\dots,(N+1)/2\}$ and $V_\mathcal{B}=\{(N+1)/2+1,\dots,N\}$ for $N$ odd or $V_\mathcal{A}=\{2,\dots,N/2+1\}$ and $V_\mathcal{B}=\{N/2+1,\dots,N\}$ for $N$ even. Finally, the associated subgraphs  $G_\mathcal{A}(V_\mathcal{A},E_\mathcal{A})$ and $G_\mathcal{B}(V_\mathcal{B},E_\mathcal{B})$ are fixed to be complete subgraphs (\emph{i.e.} all pairs of vertices in $V_\mathcal{A}$ are connected, and similarly for $V_\mathcal{B}$).

\begin{figure}[tb]
      \centering
      \includegraphics[width=5cm]{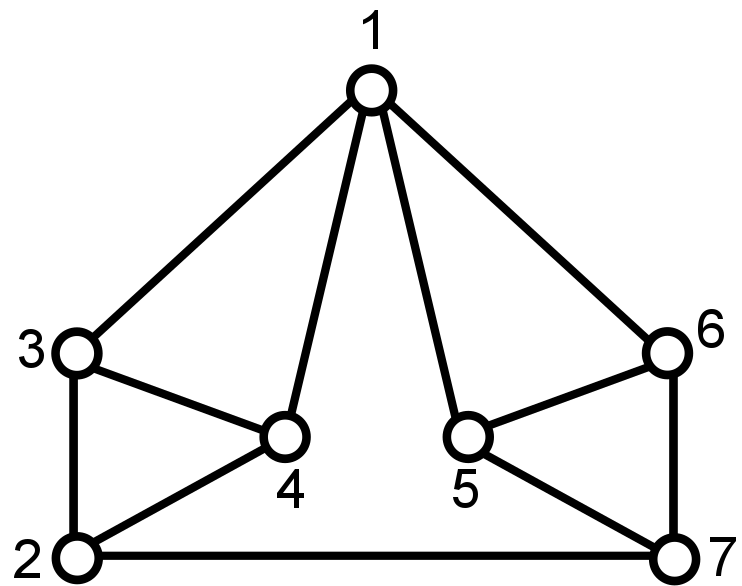}
      \caption{An example of our graph construction for odd $N$ with $N=7$. Here the two complete subgraphs are the two triangles \{2,3,4\} and \{5,6,7\}; $\alpha(G)=2$.}
			\label{fig:N7}
\end{figure}

\begin{figure}[tb]
      \centering
      \includegraphics[width=5cm]{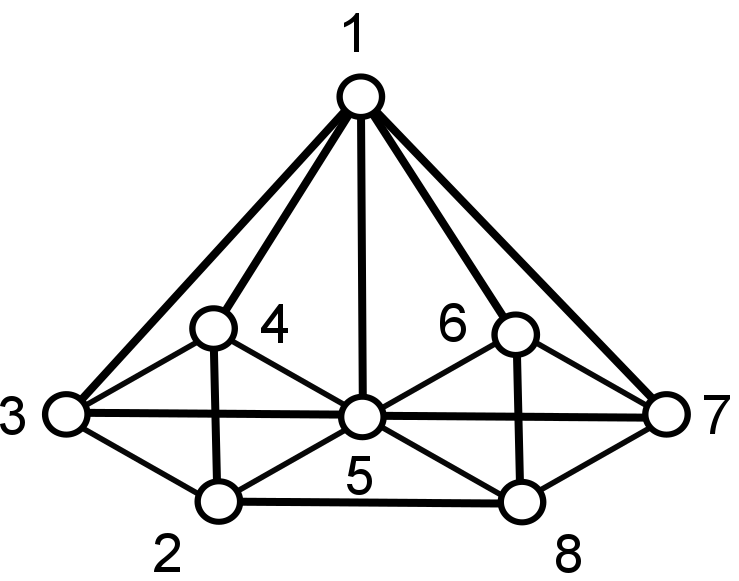}
      \caption{An example of our graph construction for even $N$ with $N=8$. The two complete subgraphs \{2,3,4,5\} and \{5,6,7,8\} have four vertices each with one common vertex \{5\}; $\alpha(G)=2$.}
			\label{fig:N8}
\end{figure}

Fig. \ref{fig:N7} shows an example of the odd case with $N=7$. We can see the two complete graphs composed by two triangles. Fig. \ref{fig:N8} shows an example of the even case with $N=8$, where we can also see the two complete subgraphs which share one vertex.

\subsection{Hardy-like paradox}
\subsubsection{Formulation of the paradox}


The exclusivity condition for the graphs defined above imposes the following conditions on the outcomes: for all vertices $i\in V$ and $j\in V$ that are connected, the probability that $X_i=1$ and $X_j=1$ reads
\begin{equation}\label{eq:hp1}
P(1,1\vert i,j)=0.
\end{equation}


As with the pentagon example before, to derive a Hardy-like paradox we will impose additional conditions. In particular we insist that for each of the sets $V_\mathcal{A}$ and $V_\mathcal{B}$ there should be at least one measurement answering `yes'. That is, for some $i \in V_\mathcal{A}$, $X_i=1$ and similarly for $V_\mathcal{B}$.
In terms of joint probabilities we have:
\begin{align}\label{eq:hp2}
P(0\dots 0\vert \{i\}_{i\in V_\mathcal{A}})=0,\notag\\
P(0\dots 0\vert \{i\}_{i\in V_\mathcal{B}})=0.
\end{align}

From the conditions in Eqs.~(\ref{eq:hp1}) and (\ref{eq:hp2}) and by assuming that the measurement outcomes can be described by a deterministic non-contextual theory, \emph{i.e.} by assuming predetermined and independent outcomes to each measurement, we can conclude that:
\begin{equation}\label{eq:hp3}
P(1\vert 1)=0.
\end{equation}
This equation can be seen by assigning the values 1 or 0 to each vertex of the graph.
In particular, if we want to assign the value $X_1=1$ to vertex $1$, the only other vertices that can possibly also have the value $X_i=1$ assigned are the vertices $2$ and $N$; all the other vertices must have the assigned outcome $X_i=0$. But because the vertices $2$ and $N$ are connected, they cannot simultaneously have the outcome 1. It is then impossible to satisfy the two conditions in Eq.~(\ref{eq:hp2}). Hence, under a deterministic non-contextual theory it is not possible to have the value $1$ assigned to vertex $1$. In other words, the probability $P(1\vert 1)$ has to be equal to zero to verify the exclusivity relations implied by the graph in Eq.~(\ref{eq:hp1}) and the additional constraints in Eq.~(\ref{eq:hp2}).
Since this must be true for all determinsitic non-contextual assignments, it is true for any convex mixture of them also.

\subsubsection{Quantum contextuality}

We will now see that in quantum mechanics it is possible to satisfy conditions Eqs.~(\ref{eq:hp1}) and (\ref{eq:hp2}) and yet also satisfy $P(1 \vert 1)\neq 0$.
In order to show this for $d>3$, we construct an example of a set of vectors for the measurements and a quantum state, for all $N>5$. For a given graph of the family, the vectors and the measured state are all qudits of dimension equal to $d = N-2$. Note that this is larger than the maximal clique of the graph, namely $N/2$ in our case, which is the lower bound for $d$ as explained previously. It may be possible to close this gap with a different construction.


To satisfy the two conditions in Eq.~(\ref{eq:hp2}) the quantum state has to be a linear combination of the vectors $\{\vert v_i\rangle\}_{i\in V_\mathcal{A}}$ and a linear combination of the vectors $\{\vert v_i\rangle\}_{i \in V_\mathcal{B}}$ simultaneously. Note that since $G_\mathcal{A}$ and $G_\mathcal{B}$ are complete subgraphs, the associated vectors are orthonormal states. This condition will therefore ensure that at least one outcome is equal to 1.
In other words, the quantum state is restricted to be in the following form:
\begin{align}\label{eq:extraNpsi}
\vert \psi \rangle= \sum_{i\in V_\mathcal{A}} \alpha_{i} \vert v_{i}\rangle,\notag\\
\vert \psi \rangle= \sum_{i\in V_\mathcal{B}} \beta_{i} \vert v_{i}\rangle,
\end{align}
where $\alpha_{i}$ and $\beta_{i}$ are complex numbers.


We can now provide different constructions depending on the parity of $N$.

\paragraph*{For $N$ odd.}We impose:
\begin{align}\label{eq:psiv1Nodd}
\vert \psi \rangle= \frac{1}{\sqrt{3}} (\vert 0\rangle + \vert \frac{N-3}{2}\rangle + \vert N-3\rangle),\notag\\
\vert v_1 \rangle= \frac{1}{\sqrt{3}} (\vert 0\rangle - \vert \frac{N-3}{2}\rangle + \vert N-3\rangle),
\end{align}
where $\{\vert 0\rangle,\vert 1\rangle , ..., \vert N-3\rangle \}$ is an arbitrary basis of the $d$-dimensional Hilbert space. Then,
\begin{align}\label{eq:P11Nodd}
P(1\vert 1)&= \vert \langle v_1 \vert \psi \rangle \vert ^2\notag\\
&= \frac{1}{9}.
\end{align}
This ensures a constant value for $P(1\vert1)$ for all $N$.

In order to construct the vectors of $V_\mathcal{A}$ and $V_\mathcal{B}$, we now define the operator $\mathcal{X}=\sum_{i=0}^{N-3}\vert i \rangle\langle N-3-i\vert$ and we choose to assign to each vector of the set $V_\mathcal{A}$ a vector in the set $V_\mathcal{B}$: $\forall i \in V_\mathcal{A}$, $\exists j \in V_\mathcal{B}$ s.t. $\mathcal{X}\vert v_i\rangle=\vert v_j\rangle$.
For example, $\mathcal{X}\vert v_2\rangle=\vert v_N\rangle$. Hence, it is enough to verify that the state is decomposable in one of the sets (either $V_\mathcal{A}$ or $V_\mathcal{B}$) because the quantum state $\vert \psi \rangle$ is invariant under $\mathcal{X}$. From the conditions imposed by the graph, each vector of the set $V_\mathcal{A}$ or $V_\mathcal{B}$ is orthogonal to all the other vectors of the same set but is not orthogonal to any vector of the other set; except for $\vert v_2\rangle$ and $\vert v_N\rangle$, which are required to be also orthogonal to each other.

We now set
\begin{align}\label{eq:v2vNNodd}
\vert v_2 \rangle= \vert 0\rangle,\notag\\
\vert v_N \rangle= \vert N-3\rangle.
\end{align}
This makes these two vectors not only orthogonal to each other but also non-orthogonal to $\vert v_1\rangle$.
All vectors $\{\vert v_i\rangle\}$, $i\in V_\mathcal{A}$, $i\neq 2$ are of the following form:
\begin{equation}\label{eq:viNodd}
\vert v_i\rangle= \vert \frac{N-3}{2}\rangle + \sum_{k=\frac{N-1}{2}}^{N-4} c_{i,k}\vert k\rangle +\vert N-3\rangle.\\
\end{equation}
For simplicity we do not include normalisation here or in some of what follows, however where this is the case normalisation plays no important role.

The associated vector $\vert v_j\rangle$ in the set $V_\mathcal{B}$ of a vector $\vert v_i\rangle$ in $V_\mathcal{A}$ is:
\begin{align}
\vert v_j\rangle&= \mathcal{X}\vert v_i\rangle\notag\\
 \vert v_j\rangle&= \mathcal{X}\vert \frac{N-3}{2}\rangle + \sum_{k=\frac{N-1}{2}}^{N-4} c_{i,k}\mathcal{X}\vert k\rangle +\mathcal{X}\vert N-3\rangle\notag\\
 \vert v_j\rangle&= \vert \frac{N-3}{2}\rangle + \sum_{\ell=1}^{\frac{N-5}{2}} c_{i,N-3-\ell}\vert \ell\rangle +\vert 0\rangle.
\end{align}

If we take one vector $\vert v_{i\neq 2}\rangle$ in $V_\mathcal{A}$ and one vector $\vert v_{j\neq N}\rangle$ in $V_\mathcal{B}$ , the inner product between the two vectors is: $\langle v_{i\neq 2}\vert v_{j\neq N}\rangle = 1$. Hence, we verify the (non) orthogonality conditions that the graph demands.

We now need to verify the orthogonality within each set (see Eq.~(\ref{eq:hp2})) and also that one set can generate the quantum state $\vert \psi \rangle$. The coefficients $\{c_{i,k}\}$ in Eq.~(\ref{eq:viNodd}) can be tailored to verify these two conditions. In fact, this problem can be solved using a matrix formulation: let $\mathcal{M}$ be a $\frac{N-3}{2}\times \frac{N-5}{2}$ matrix, for which each row corresponds to one of the $\frac{N-3}{2}$ vectors $\vert v_{i\neq 2}\rangle \in V_\mathcal{A}$ and the elements are equal to the coefficients $\{c_{i,k}\}$. Then, 
\begin{enumerate}[(i)]
\item The orthogonality within each set requires the inner product between two rows to be equal to $-2$.
\item The possibility to generate the quantum state $\vert \psi \rangle$ of Eq.~(\ref{eq:psiv1Nodd}) requires the sum of the coefficients of each column to be equal to zero.
\end{enumerate}

In other words:
\begin{enumerate}[(i)]
\item $\forall$ $(i,j)$ with $i\neq j$, $\sum_k c_{i,k} c_{j,k} = -2$.
\item $\forall$ $j$, $\sum_i c_{i,j} = 0$.
\end{enumerate}

It can be proven by induction that such a matrix exists for odd $N\geq 7$.

For example, for $N=7$:
\begin{equation*}
   \mathcal{M}_{2\times 1} =
   \begin{pmatrix}
       -\sqrt{2} & \\
       \sqrt{2} & \\
   \end{pmatrix}
	,
\end{equation*}

and for $N=9$:
\begin{equation*}
   \mathcal{M}_{3\times 2} =
   \begin{pmatrix}
       -2 &0 \\
       1 & -\sqrt{3} \\
			 1 & \sqrt{3} \\
   \end{pmatrix}
	.
\end{equation*}

Then, the quantum state in Eq.~(\ref{eq:psiv1Nodd}) can be obtained by using the following linear combination:
\begin{align}\label{eq:psiNodd}
\vert \psi \rangle= \frac{2}{\sqrt{3}(N-3)}\big( \sum_{i\in V_\mathcal{A},i\neq 2}\vert v_i\rangle +\frac{N-3}{2}\vert v_2\rangle\big),\notag\\
\end{align}
and hence, it is possible to obtain, as in Eq.~(\ref{eq:P11Nodd}), $P(1\vert 1)= \frac{1}{9}$.

\paragraph*{For $N$ even.} In this case, the sets $V_\mathcal{A}$ and $V_\mathcal{B}$ share one vertex with the associated vector $\vert v_{N/2+1} \rangle$. By following a similar procedure as for $N$ odd we can derive the sets of measurements and the quantum state. More specifically, the vectors $\vert v_2\rangle$ and $\vert v_N \rangle$ are the same as before, Eq.~(\ref{eq:v2vNNodd}), while $\vert \psi \rangle$ and $\vert v_1 \rangle$ are now as follows:
\begin{align}\label{eq:psiv1v2vNNeven}
\vert \psi \rangle&= \frac{1}{\sqrt{6}} (\sqrt{2}\vert 0\rangle + \vert \frac{N}{2}-2\rangle + \vert \frac{N}{2}-1\rangle + \sqrt{2}\vert N-3\rangle),\notag\\
\vert v_1 \rangle&= \frac{1}{\sqrt{6}} (\sqrt{2}\vert 0\rangle - \vert \frac{N}{2}-2\rangle - \vert \frac{N}{2}-1\rangle +\sqrt{2} \vert N-3\rangle),
\end{align}

The vector corresponding to the shared vertex of the two sets $V_\mathcal{A}$ and $V_\mathcal{B}$ is:
\begin{equation}\label{eq:vN/2Neven}
\vert v_{N/2+1} \rangle= \frac{1}{\sqrt{2}} (-\vert \frac{N}{2}-2\rangle + \vert \frac{N}{2}-1\rangle),\\
\end{equation}
and all vectors $\vert v_i\rangle$ in the set $V_\mathcal{A}$ except for $\{\vert v_2\rangle,\vert v_{N/2+1} \rangle \}$ are of the following form:
\begin{align}\label{eq:viNeven}
\vert v_i\rangle= \vert \frac{N}{2}-2\rangle + \vert \frac{N}{2}-1\rangle + \sum_{k=\frac{N}{2}}^{N-4} c_{i,k}\vert k\rangle +\sqrt{2}\vert N-3\rangle.\notag\\
\end{align}

As in the odd $N$ case, each vector in the set $V_\mathcal{B}$ can be obtained by applying $\mathcal{X}$ to one vector of the set $V_\mathcal{A}$, and similarly we can build a matrix with the coefficients $\{c_{i,k}\}$, which has to verify two properties:
\begin{enumerate}[(i)]
\item $\forall$ $(i,j)$ with $i\neq j$, $\sum_k c_{i,k} c_{j,k} = -4$.
\item $\forall$ $j$, $\sum_i c_{i,j} = 0$.
\end{enumerate}

It can be proven by induction that such a matrix exists for even $N > 6$. For $N = 6$ such a matrix is not defined as there are no $\{c_{i,k}\}$ coefficients. In this case, it is enough to take $\vert v_3 \rangle = \vert 1 \rangle + \vert 2 \rangle + \sqrt{2}\vert 3 \rangle$ and $\vert v_4 \rangle = \sqrt{2}\vert 0 \rangle + \vert 1 \rangle + \vert 2 \rangle$.

Finally, the quantum state can be obtained by preparing the linear combination:
\begin{align}\label{eq:psiNeven}
\vert \psi \rangle= \frac{2}{\sqrt{6}(N-4)}\big( \sum_{i\in V_\mathcal{A},i\neq 2,N/2+1}\vert v_i\rangle +\frac{N-3}{2}\vert v_2\rangle\big),\notag\\
\end{align}
and we obtain, as before, the value $P(1\vert 1)=\frac{1}{9}$.

\subsubsection{Majorana representation}

The family of graphs that we have built feature a high connectivity, hence allowing us to extend the Hardy paradox test of contextuality to qudits of dimension greater than three. An interesting way of visualizing the symmetry of our graph construction is to use the Majorana representation, where a $d$-dimensional system is described (up to a global phase) by $d-1$ points on the surface of a sphere \cite{majorana1932atomi,bengtsson2007geometry}. Given a state $|\psi\rangle=\sum_{k=0}^{d-1} a_k |k\rangle$, the points are found by computing the zeros $\alpha$ of the polynomial $f(\psi):=\sum_{k=0}^{d-1} \sqrt{{ d-1 \choose k}}a_k \alpha^k$, and taking their stereographic projection. That is for a zero $\alpha_i$ one associates a point at angles $\theta_i, \phi_i$ with $\alpha_i = e^{-i \phi_i} \tan\left(\frac{\theta_i}{2}\right)$ along with the convention that a zero solution is identified with the north pole and a polynomial $f(\psi)$ of order $k<d-1$ has $d-1-k$ points at the south pole.

The Majorana representation has a plethora of applications including quantum chaos \cite{leboeuf1991phase}, Berry phase \cite{hannay1998berry}, classicality \cite{zimba2006anticoherent}, many-body physics \cite{ribeiro2007thermodynamical}, and in proofs of contextuality \cite{penrose2000bell}.
Using the well known map between a single $d$-dimensional system and $d-1$ spin-$1/2$ systems restricted to the permutation symmetric states, the Majorana representation has also been used to study entanglement \cite{markham2011entanglement,AMM:njp10}, symmetry and non-locality \cite{WM:prl12}. By illustrating our states and measurement bases in this representation we may thus see connections between these diverse topics.

\begin{figure}[tb]
      \centering
      \includegraphics[width=5.5cm]{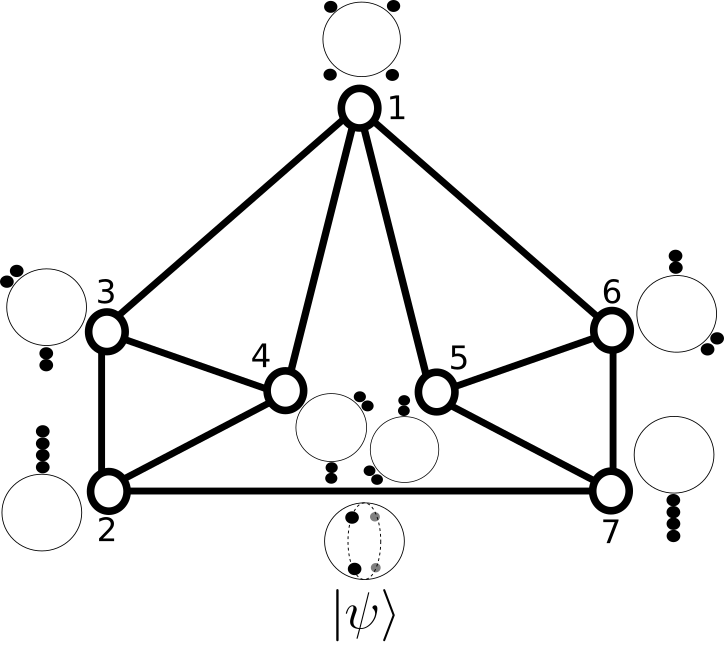}
      \caption{$N=7$ corresponding to $d=5$; the quantum state and measurement vectors are represented by 4 Majorana points.}
			\label{fig:MajN7}
\end{figure}

\begin{figure}[tb]
      \centering
      \includegraphics[width=5.5cm]{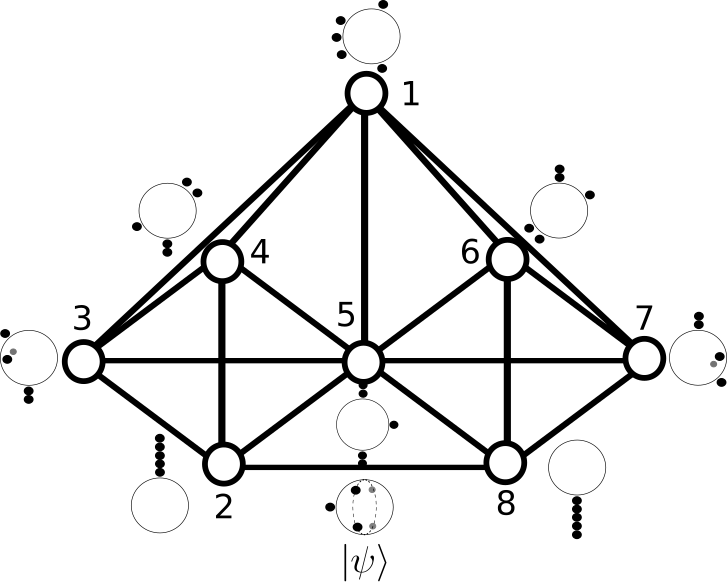}
      \caption{$N=8$ corresponding to $d=6$; the quantum state and measurement vectors are represented by 5 Majorana points.}
			\label{fig:MajN8}
\end{figure}

In particular in our construction two interesting properties can be observed. First, the symmetry in the construction corresponds to a symmetry of the points - the operator $\mathcal{X}$ acts as a flip in the $X$ axis, so that the point distributions of $|v_1\rangle$ and the state $|\psi\rangle$ have symmetry about an $X$ flip, and the states in $V_{\mathcal{A}}$ and $V_{\mathcal{B}}$ are related by $X$ flips. The second property is that the construction contains a lot of degeneracy (\emph{i.e.} points sitting on top of each other), which was recently discovered to offer different types of entanglement and non-locality in the multiparty scenario \cite{WM:prl12,wang2013nonlocality}. In Figs. \ref{fig:MajN7} and \ref{fig:MajN8} we depict the Majorana points of the quantum state and the measurement vectors for examples of odd ($N=7$) and even ($N=8$) numbers of vertices, respectively, where the big circles are a schematic representation of the Bloch sphere and the black dots give the positions of the Majorana points of the corresponding vectors.



\subsection{Extension of the KCBS Inequality}

\subsubsection{Classical bound}

Different extensions of the KCBS inequality using a graphical approach have been studied in the past \cite{AQBCC:pra13,CBCB:prl13} and even experimentally implemented \cite{ACGBXLABSC:pra15}. Here we are interested in providing an extension tailored to our construction.

The extended KCBS inequality is given in Eq.~(\ref{eq:beta}). The classical bound is obtained by the independence number of the graph, $\alpha(G)$, which is in fact equal to the sum of all the outcomes of the vertices. As we have explained previously, because of the exclusivity relation assumed in building our family of graphs, if the outcome 1 is assigned to any vertex, only one other vertex may possibly have the outcome 1. Hence, $\alpha(G)=2$ for all graphs $N > 5$ giving the inequality:
\begin{align}\label{eq:kcbsN}
\beta=\sum_{i=1}^N \langle X_i \rangle  \leq 2.
\end{align}

\subsubsection{Quantum violation}

We now calculate the quantum violation for our construction. Because $V_\mathcal{A}$ and $V_\mathcal{B}$ are both orthogonal sets which generate $\vert \psi \rangle$, we can write:
\begin{align}
\sum_{i\in V_{\mathcal{A}}} \vert \langle v_i \vert \psi \rangle \vert ^2 = \sum_{i\in V_{\mathcal{B}}} \vert \langle v_i \vert \psi \rangle \vert ^2=1.
\end{align}

The value of $\beta$ for the set of vectors $\{\vert v_i\rangle\}$ and the quantum state $\vert \psi \rangle$ for $N$ odd is:
\begin{align}\label{eq:kcbsNodd}
\beta & = \vert \langle v_1 \vert \psi \rangle \vert ^2 + \sum_{i\in V_{\mathcal{A}}} \vert \langle v_i \vert \psi \rangle \vert ^2 + \sum_{i\in V_{\mathcal{B}}} \vert \langle v_i \vert \psi \rangle \vert ^2\notag\\
& = \frac{1}{9} + 2.
\end{align}

For even values of $N$, the state $\vert \psi \rangle$ in Eq.~(\ref{eq:psiNeven}) has no contribution in $\vert v_{N/2+1} \rangle$. Hence, $\vert \langle v_{N/2+1} \vert \psi \rangle \vert ^2 = 0$. The value of $\beta$ in this case is:
\begin{align}\label{eq:kcbsNeven}
\beta & = \vert \langle v_1 \vert \psi \rangle \vert ^2 + \!\!\sum_{i\in V_{\mathcal{A}},i\neq N/2+1} \!\!\!\!\!\vert \langle v_i \vert \psi \rangle \vert ^2\notag\\
      & +\!\! \sum_{i\in V_{\mathcal{B}},i\neq N/2+1} \!\!\!\!\!\vert \langle v_i \vert \psi \rangle \vert ^2\notag\\
      & = \frac{1}{9} + 2.
\end{align}

As in the case of the pentagon, we expect that the state and measurements that maximally violate the inequality will not be those satisfying the paradox. Unfortunately searching over both measurement settings and states quickly becomes too difficult numerically. However we were able to search over states, using the same measurement settings as those used for the paradox, and we find indeed a higher violation of around $2.22$ up to $d=10$.
A better violation may be obtained by extending this search also over the measurements.

\section{Discussion}

The experimental verification of contextuality is fraught with difficulty (see for example \cite{BK04,Spekkens14} and references therein). One crucial issue is the problem of imprecision and errors associated to any physical implementation of a measurement. In particular, it is not in practice possible to be sure that measurements in different contexts are really the same - we can only try to measure \emph{almost} the same thing. For example, 
the position of a polarizing plate cannot be guaranteed to be exactly the same for successive measurements, although the drift may be very small. This is especially relevant if other intermediate realignments must be made to change context.
In terms of our graph construction, looking at the pentagon, Fig.~\ref{fig:N5}, for example, $X_1$ should be measured with $X_2$ in one context and with $X_5$ in another. In quantum mechanics, these measurements are described by the projectors $P_i$, so in both contexts we should be measuring $P_1$. However in reality in the different contexts there would be some small difference no matter how hard we tried and so we would instead measure some $P_1$ with $P_2$ and a slightly different $P_1'$ with $P_5$, where $P_1$, $P_1'$ might be associated to very close but not exactly the same angles of polarisation. Strictly speaking we should then associate this to a distinct measurement when comparing with what can be done classically. In a series of earlier works \cite{Meyer:prl99,Kent:prl99,CK:pra2000}, it was shown that it is always possible to find sufficiently close projectors to make a contextual model for any non-perfect precision quantum mechanical measurement (see also \cite{BK04}).

There are several approaches to addressing this issue, the main idea being that we would like somehow to say that if the measurements are close, then any contextual hidden variable model describing them are close too. We briefly consider the approach laid out in \cite{Winter:jpa14}, which is easily amenable to our constructions. There, the notion of `ontological faithfulness' is introduced, which puts a statement on the closeness of the measurement statistics in different contexts. That is, a model is said to be `$\epsilon$-ontologically faithful non-contextual' ($\epsilon$-ONC) if the probability of results differing for measurements associated to the same vertex in different contexts is less than or equal to $\epsilon$. Furthermore, for a graph of $N$ vertices, when each vertex is involved in only two contexts (as is the case for our examples), and for associated inequalities of the form of Eq.~(\ref{eq:kcbsN}), denoting as $\Delta$ the difference between the observed violation and the classical bound, if
\begin{align}
\epsilon & < \frac{\Delta}{N}, \text{ for $N$ odd}, \notag\\
\epsilon & < \frac{\Delta}{N+3}, \text{ for $N$ even},
\end{align}
then there is no ontologically faithful non-contextual model which matches the results \cite{Winter:jpa14}.
For our constructions we have $\Delta=\frac{1}{9}$, so if we can achieve a precision of $\epsilon < \frac{1}{9 N}$ (for $N$ odd) and $\epsilon < \frac{1}{9(N+3)}$ (for $N$ even) we can be sure there is no $\epsilon$-ONC hidden variable model achieving our results. We see then that naturally this gets more difficult to ensure as $N$ becomes larger.

We note here that there are several alternative approaches, notably the ones that do not require additional definitions like ONC, as they are rather built from an operational definition of contextuality \cite{Pusey:arxiv15,MOKRS:arxiv15}.

\section*{CONCLUSION}

We have developed a family of graphs for which we gave a logical and an inequality-based proof of contextuality in the same conditions. Our construction appears to be a natural extension scenario for the Hardy-like paradox proof of contextuality. We have provided an explicit way to obtain the measurement settings and the quantum state to achieve an experimental proof of contextuality for an arbitrary dimension qudit. We have also proved that within the condition of the paradox, the violation of the generalization of the KCBS inequality for all qudits is at least the same as for the qutrit. An open question is whether alternative graph constructions exist that may lead to better violations for qudits. Proposing practical ways of demonstrating contextuality for qudits in our framework, in the line for instance of \cite{BJMC:prl14}, is also a challenging and interesting subject for further research.

\noindent {\bf Acknowledgements.}
We acknowledge support from the ANR project COMB and the ville de Paris project CiQWii.

\bibliography{biblio}

\end{document}